\documentclass[twocolumn]{aastex62}

\shorttitle{K2, TESS, and {\it Spitzer} transits}
\shortauthors{Duck et al. 2021}

\begin{document}
\title{K2, {\it Spitzer}, and TESS Transits of Four Sub-Neptune Exoplanets}

\author[0000-0002-4531-6899]{Alison Duck}
\affiliation{Department of Astronomy, University of Maryland, College Park, MD 20742, USA}
\affiliation{Department of Astronomy, The Ohio State University, 140 West 18th Avenue, Columbus, OH 43210, USA}

\author[0000-0001-5737-1687]{Caleb K. Harada}
\altaffiliation{NSF Graduate Research Fellow}
\affiliation{Department of Astronomy, University of California Berkeley, University Drive, Berkeley, CA 94720, USA}
\affiliation{Department of Astronomy, University of Maryland, College Park, MD 20742, USA}

\author[0000-0001-6771-4583]{Justin Harrell}
\affiliation{Department of Astronomy, University of Maryland, College Park, MD 20742, USA}
\affiliation{Department of Physics and Astronomy, University of Delaware, Newark, DE 19716, USA}

\author[0000-0002-4265-3574]{Ryan R. A. Morris}
\affiliation{Department of Astronomy, University of Maryland, College Park, MD 20742, USA}
\affiliation{School of Physics, University of New South Wales, Sydney NSW 2052, Australia}

\author{Edward Williams}
\affiliation{Department of Astronomy, University of Maryland, College Park, MD 20742, USA}
\affiliation{Department of Geology, University of Maryland, College Park, MD 20742, USA}

\author{Ian Crossfield}
\affiliation{Department of Physics, and Kavli Institute for Astrophysics and Space Research, Massachusetts Institute of Technology, Cambridge, MA 02139, USA}
\affiliation{Department of Physics and Astronomy, The University of Kansas, Lawrence, KS 66045, USA}

\author{Michael Werner}
\affiliation{Jet Propulsion Laboratory, California Institute of Technology, Pasadena, CA 91109}

\author{Drake Deming}
\affiliation{Department of Astronomy, University of Maryland, College Park, MD 20742, USA}
\affiliation{NASA Astrobiology Institute's Virtual Planetary Laboratory}

\begin{abstract}

We present new {\it Spitzer} transit observations of four K2 transiting sub-Neptunes: K2-36c, K2-79b, K2-167b, and K2-212b. We derive updated orbital ephemerides and radii for these planets based on a joint analysis of the {\it Spitzer}, TESS, and K2 photometry. We use the \texttt{EVEREST} pipeline to provide improved K2 photometry, by detrending instrumental noise and K2's pointing jitter.  We used a pixel level decorrelation method on the {\it Spitzer} observations to reduce instrumental systematic effects.  We modeled the effect of possible blended eclipsing binaries, seeking to validate these planets via the achromaticity of the transits (K2 versus {\it Spitzer}).  However, we find that {\it Spitzer's} signal-to-noise ratio for these small planets is insufficient to validate them via achromaticity. Nevertheless, by jointly fitting radii between K2 and {\it Spitzer} observations, we were able to independently confirm the K2 radius measurements.  Due to the long time baseline between the K2 and {\it Spitzer} observations, we were also able to increase the precision of the orbital periods compared to K2 observations alone.  The improvement is a factor of 3 for K2-36c, and more than an order of magnitude for the remaining planets. Considering possible JWST observations in 1/2023, previous $1\sigma$ uncertainties in transit times for these planets range from 74 to 434 minutes, but we have reduced them to the range of 8 to 23 minutes.

\end{abstract}


\section{Introduction} \label{sec:intro}

Extrasolar planets with radii as small as $1.6\,R_{\oplus}$ are capable of retaining envelopes of hydrogen and helium \citep{rogers2015}.  Results from the Kepler mission demonstrated that these sub-Neptune exoplanets are very common in our Galaxy \citep{fressin2013}, although none are present in our Solar System.  Transit spectroscopy of sub-Neptunes can probe their atmospheric composition, and thereby shed light on the nature of these worlds \citep{piette2020}.  The \textit{James Webb Space Telescope} (JWST) will be a powerful facility for transit spectroscopy of sub-Neptunes \citep{greene2016}.  However, JWST transit observations require accurate orbital ephemerides in order to plan the observations.  Also, improved planetary radii can help to estimate the signal-to-noise for transit spectroscopy, e.g., using the Transit Spectroscopy Metric \citep{kempton}.  In this paper, we report improved orbital ephemerides and radii for four sub-Neptunes that are potentially important targets for atmospheric characterization using JWST.

After the end of the prime {\it Kepler} mission, NASA's K2 mission continued to discover transiting sub-Neptunes orbiting stars near the ecliptic plane \citep{2014PASP..126..398H}. However, K2 observed each of its campaign fields for approximately 80 days, and that limited the precision of the orbital periods derived from K2 data.  As time passed, errors in transit ephemerides accumulated, making predictions of future transits difficult or impractical \citep{2016ApJ...822...39B}. Follow-up observations by JWST may be problematic for planets whose transit times are uncertain by an amount exceeding the transit (or secondary eclipse) duration \citep{2017ApJ...834..187B}.

By following K2 transit observations with {\it Spitzer} measurements of single transits over a longer span of time, the precision of orbital periods can be greatly improved, and other information such as planetary radii can be validated \citep{2017ApJ...834..187B, 2018arXiv180110177C, kosiarek2019, livingston2019}. Because {\it Spitzer} observations are only minimally influenced by stellar limb darkening \citep{2012AA...546A..14C}, a {\it Spitzer} transit exhibits a sharp ingress and egress that facilitate deriving precise transit times.  Whereas the 30 minute cadence of K2 broadens the observed ingress and egress, {\it Spitzer's} cadence of 0.4 seconds fully preserves the intrinsically sharp ingress and egress.   Moreover, the reduced limb darkening enables deriving planet radii with less dependence on the properties of the host stars, and in any case can yield independent checks on the K2 radii. 

Observations by both {\it Spitzer} and K2 can potentially help to distinguish true exoplanet transits from false-positives due to blended eclipsing binaries (BEBs). BEB eclipse depths can be strongly wavelength-dependent \citep{2015ApJ...804...59D}, whereas the depth of an exoplanet transit should not vary greatly based on the wavelength of observation\footnote{Not counting the subtle wavelength-dependent variations in transit radius due to absorption in the exoplanetary atmosphere. These variations are typically only of order ~ $\frac{2R_{p}H}{R_*^2}$, where $H$ is the scale height of the exoplanetary atmosphere, and $R_p$ and $R_*$ are the planetary and stellar radii, respectively.  This effect is thus much smaller than the chromaticity due to typical BEBs}. By observing the transit in two very different wavelengths we could in principle verify that the transit is achromatic. 

We report {\it Spitzer} transits of four K2 sub-Neptunes (K2-79b, K2-36c, K2-167b, K2-212b). Additionally, we include TESS (Transiting Exoplanet Survey Satelite) transits for K2-167b \citep{2015JATIS...1a4003R}. We improve the orbital ephemerides of these four systems, as well as provide improved measurements of their radii. We also use a Galactic model of background BEBs to determine whether achromaticity of the transits comparing K2 to {\it Spitzer} can increase the confidence in the validation of the planets.  Our targets are a small subset of K2 planets observed by {\it Spitzer} in program 11026 \citep{SpitzerProgram}. The program was designed to observe small planets orbiting stars that are bright in the infrared so that there would be an optimum signal to noise ratio in the {\it Spitzer} observations. 

Section~\ref{sec:host_stars} discusses the properties of the host stars, and in Section \ref{sec:observations}, we describe our observations from the K2 and {\it Spitzer} campaigns.  Section~\ref{sec:overview} gives an overview of our procedure for fitting the transits from the different missions. In Section \ref{sec:photometry}, we explain the photometric procedures we used to produce transit light curves.  Section \ref{sec:analysis} describes both independent and joint analyses of the K2, {\it Spitzer}, and TESS light curves. We give our results for radii and transit times in Section \ref{sec:results}. Section \ref{sec:vetting} investigates to what degree BEB false-positives can be rejected based on our K2 and {\it Spitzer} radii.  Section \ref{sec:discussion} summarizes, and Section \ref{sec:conclusion} gives some remarks on future work.

\startlongtable
\begin{deluxetable*}{lcccc}
\tablecaption{Target System Parameters From Literature \label{tab:system_params}}
\tablehead{
\colhead{} & \colhead{EPIC 201713348} & \colhead{EPIC 210402237} & \colhead{EPIC 220321605} & \colhead{EPIC 205904628}
}
\startdata
\textsl{Planet Parameters}\tablenotemark{a}	&	&	&	&	\\
K2 ID		&	\textbf{K2-36c}		&	\textbf{K2-79b}			&	\textbf{K2-212b} &	\textbf{K2-167b}		\\
$R_p/R_*$	&	$0.0323\pm0.0028$		&	$0.0261\pm0.0012$			&	$0.0367\pm0.0019$		&	$0.014\pm0.0013$		\\
$Per$ [d]	&	$5.34072\pm0.00011$	&	$10.99573\pm0.00070$	&	$9.79540\pm0.00029$		&	$9.9775\pm0.001$		\\
$a$ [AU]	&	$0.05574\pm0.00046$	&	$0.0987\pm0.0012$		&	$0.07648\pm0.00082$		&	$0.0913\pm0.0036$		\\
\textsl{Stellar Parameters}\tablenotemark{a}	&	&	&	&	\\
Kp [mag]				&	$11.531$			&	$11.801$				&	$12.588$		&	$8.220$		\\
$R_*$ [$R_\odot$]		&	$0.726^{+0.052}_{-0.048}$	& $1.247^{+0.077}_{-0.072}$	&		$0.678^{+0.052}_{-0.047}$		&		$1.499^{+0.088}_{-0.079}$		\\
$M_*$ [$M_\odot$]		&	$0.8100\pm0.0200$	&	$1.0600\pm0.0400$		&	$0.622\pm0.02$		&	$1.02\pm0.12$		\\
$T_\textrm{eff}$ [K]			&	$4953\pm71$			&	$5926\pm86$				&	$4349.0\pm50.0$		&	$5908.0\pm50.0$		\\
$\log g$ [cm/s$^2$]		&	$4.5980\pm0.01700$	&	$4.2490\pm0.07300$		&	$4.72\pm0.1 $		&	$3.88\pm0.1 $		\\
$(\mu_1,\mu_2)$ (K2)\tablenotemark{b}
			&	$(0.6147,0.1061)$	&	$(0.4899,0.1809)$	&	$(0.6670,0.0790)$		&	$(0.4899,0.1809)$		\\
$(\mu_1,\mu_2)$ ({\it Spitzer})\tablenotemark{b}
			&	$(0.0965,0.1144)$	&	$(0.0774,0.1029)$	&	$(0.0774,0.1029)$&	$(0.0774,0.1029)$		\\
\enddata
\tablenotetext{a}{Planetary and stellar parameters for K2-36c and K2-79b are from \citet{crossfield}, except for the stellar radii that are from \citet{hardegree2020}.
Planetary and stellar parameters for K2-212b and K2-167b are from \citet{mayo}, except for the stellar radii, that are from \citet{hardegree2020}.}
\tablenotetext{b}{Quadratic limb darkening coefficients for {\it Spitzer} 4.5\,$\mu$m and {\it Kepler} bandpasses from \citet{2012AA...546A..14C} and \citet{2013AA...552A..16C}.}
\end{deluxetable*}

\section{Properties of the Host Stars} \label{sec:host_stars}
 Properties of the stars and their environment on the plane of the sky can affect our analysis. Basic data for the host stars are listed in Table \ref{tab:system_params}. We investigated their possible photometric variability (e.g., due to rotational modulation of star spots), and the possible presence of nearby stars within the same K2 pixels, that could dilute the transits.  

We determined that K2-79, K2-167, and K2-212 are not significantly variable in the K2 data.  Over periods of 64, 32, and 78 days, the K2 flux of those three stars varied by 1.7\%, 0.1\%, and 0.8\%, respectively.  The variations have no obvious periodic behavior, being gradual changes over the entire K2 interval of visibility.  K2-167 was also observed by TESS; during two intervals of 27 and 25 days, we find variations of 0.30\% and 0.14\%, respectively, and the variations are gradual, not periodic.  There will be no effect on our transit analysis, because: 1) the photometric variations are small, 2) transit analyses are based on relative changes, not being sensitive to the absolute flux, and 3) the variations do not indicate the presence of star spots.  In the case of K2-36, we find variations of $\pm0.7$\% with a period of $\sim$\,10 days.  That suggests rotation of a weakly-spotted star, but the amplitude is too small to produce significant bias in our transit analysis.

Since K2 pixels cover $4 \times 4$ arcsec, the flux from nearby stars can be included and can potentially dilute the transit depths derived by our fitting procedure.  We consulted the literature of high resolution imaging to search for companion stars in transiting systems \citep{ngo2015, ngo2016, wollert2015a, wollert2015b, evans2018, ziegler2018}, but we do not find any of our host stars in those investigations.  Therefore, we searched for nearby optical companions that might dilute the K2 transits using images from the Sloan Digital Sky Survey, the 2MASS survey, and the Gaia EDR3 catalog \citep{gaia2020}, for the regions covered by the K2 pixels used in our analyses.  We find no other stars included in the K2 photometry that are bright enough to produce significant dilution of the transits.

Following conventional practice, we measure transit depths relative to the flux from the host star, yielding the ratio of planetary to stellar radii.  We obtain the exoplanetary radii by multiplying that ratio by the stellar radius, using the stellar values from \citet{hardegree2020}.

\section{Observations} \label{sec:observations}

For the K2 photometry, we used the observations at 30 minute cadence from campaigns 1, 3, 4, and 8.  Rather than use the aperture photometry available in the MAST archive, we elected to extract improved photometry for each planetary system using the \texttt{EVEREST} (EPIC Variability Extraction and Removal for Exoplanet Science Targets,~\citealp{2016AJ....152..100L}) python package, as explained in Sec.~\ref{sec:photometry}.  We also observed all four planets spanning the predicted times of transit using {\it Spitzer} at 4.5\,$\mu$m, with 0.4-second exposures in subarray mode, planned by program 11026 \citep{SpitzerProgram}. In the case of K2-167 we also included observations from 2018 and 2020 TESS observing campaigns.

\section{Overview of the Analysis}\label{sec:overview}

We analyzed transits from three separate space missions (K2, {\it Spitzer}, and TESS).  Prior to describing the analysis of the transits from each mission, and then a joint analysis, we here describe the overall process so that our procedures will be clear.  Details are given in following sections.

We first analyzed the data from each mission separately.  Then we combined them in a joint analysis.  We have K2 and Spitzer transits for all four planets, and TESS transits for only K2-167.  For each planet we began by using the \texttt{EVEREST} package to produce K2 photometry, and we phased the multiple K2 transits to a common orbital phase scale.  We then fit to the phased K2 photometry to derive the radius and orbital parameters of each planet.  As part of that analysis, we solve for an improved orbital period that minimizes the scatter in the phased transit.  We note that K2-36 also hosts a second planet (K2-36b) with an orbital period of 1.42 days \citep{mayo}.  We verified that transits of the 'b' planet do not occur sufficiently close to the 'c' transits to affect our analysis.

Next, we extracted {\it Spitzer} photometry and fit to the {\it Spitzer} transit; we have one {\it Spitzer} transit for each planet.  We solved for the transit depth (i.e., ratio of planetary to stellar radius) and central time of the transit in the {\it Spitzer} bandpass (4.5\,$\mu$m).  For both the K2 and {\it Spitzer} analyses, we adopted quadratic limb darkening coefficients based on model atmospheres, and we froze those coefficients during the fitting process.

With the K2 and {\it Spitzer} analyses done independently, we use those individual transits and estimates of the orbital parameters in a joint fitting process.  That joint fitting refines the orbital parameters, including the period.  Our planetary radii are based on fitting for $R_p/R_s$, and converting to the planetary radius using stellar radii from \citet{hardegree2020}.  For K2-167 we also phase the TESS transits to a common orbital phase scale using the period derived in the K2/{\it Spitzer} analysis, and again using model atmosphere quadratic limb darkening coefficients.  The noise level of the TESS data is higher than the noise level of the K2 data, so we use the TESS transits only for their transit times, to extend the baseline for determining the orbital period.

With the above analyses done, we adopt a final set of orbital parameters, and then re-fit each individual transit to determine a refined transit time.  We use those individual transit times in a least-squares fit for our final determination of the orbital periods.

\begin{figure*}
    \centering
    \includegraphics[width=6in]{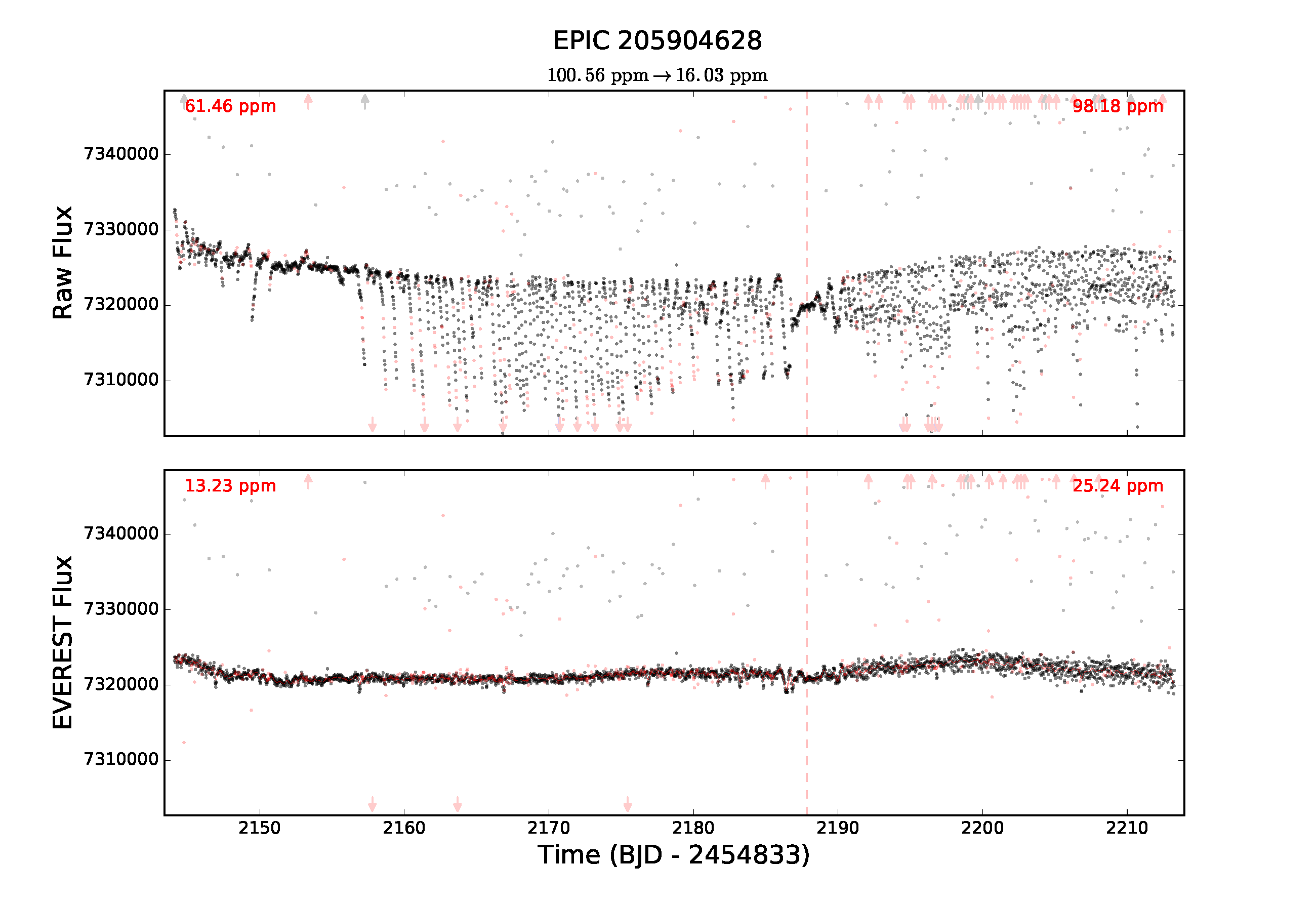}
    \caption{Raw light curve (top) and \texttt{EVEREST} de-trended light curve (bottom) for K2-167b. Note the large amplitude noise due to pointing excursions in the raw light curve, that are successfully corrected by \texttt{EVEREST}. The dashed red line indicates a break in the observation.  The arrows indicate removed outlier points that extend past the the plotted range. The other red points are outliers identified by \texttt{EVEREST} that have been removed. Thruster firing contribute to the collection of high flux outliers. The red numbers in each section show the 6h Combined Differential Photometric Precision (CDPP) for each side of the break. We see that the the CDPP drops from the raw 100 pm to about 16 ppm, indicating efficiently de-trended photometry.}
    \label{fig:everest}
\end{figure*}

\section{Photometry} \label{sec:photometry}
\subsection{{\it Spitzer}~Photometry}

 We performed aperture photometry on the {\it Spitzer} images as described by \citet{deming15} and \citet{garhart} and summarized here. Our photometry used numerical apertures with 11 constant radii, which we varied from 1.8 to 3.5 pixels in increments of approximately 0.2 pixels.  We also used numerical apertures with variable radii based on the noise pixel formulation \citep{lewis13}, with 11 constants added to the noise pixel radius.  We varied those constants from 0 to 1.2 pixels in increments of 0.1 pixel.  Both a 2-D Gaussian fit, and a center-of-light calculation served to determine the centroid of the stellar image, and we used a Gaussian fit to a histogram of pixels not containing the star in order to define and subtract the sky and instrument background.  This process produces four sets of photometry (2 aperture types, each using 2 centroiding methods), and each set of photometry contains fluxes based on 11 apertures.

In order to remove {\it Spitzer's} well known intra-pixel sensitivity variations, we used Pixel-Level Decorrelation (PLD, \citealp{deming15}).  Our updated implementation of PLD (described by \citealp{garhart}) used 12 pixels surrounding the stellar centroid in a 4x4 pattern without the corners.  For each of the four sets of photometry, the fitting code bins the pixel coefficients and the photometry using a range of bin sizes, and the code selects the best aperture radius and bin size that produces the best results in the Allan deviation relation (see \citealp{garhart} for details).  We re-fit each set of photometry using both linear and quadratic temporal ramps, choosing between them based on a Bayesian Information Criterion, and we choose among the four sets of photometry based on the best ratio of the decorrelated scatter to the photon noise. In doing these fits, we include the transit shape as defined by the algorithm from \citet{mandel}, with quadratic limb darkening. We verified that the transit curves from \citet{mandel} and \texttt{BATMAN} \citep{2015ascl.soft10002K}, are mutually consistent, for the same input parameters.  At this stage of the process, we fix the orbital parameters at the values in Table \ref{tab:system_params}, but we allow the transit time and radius ratio to vary in the fit. 

After the preliminary fit described above, we subtract the instrumental portion of the fitted curve, yielding time series photometry with {\it Spitzer's} intra-pixel effect removed, but with the observed transit remaining.  This preliminary fit in effect served the purpose of masking the transit and solving thoroughly for, and removing, the intra-pixel effect on the photometry.  That permits us to now fit the Spitzer and K2 transits in a joint fit, as described in Sec.~\ref{subsec:joint}.  However, we also re-fit the decorrelated {\it Spitzer} data independently of the K2 data, but using a procedure that is consistent between both missions.  We now describe that process for the K2 data.

\subsection{K2 Photometry} \label{sec:K2phot}

We use \texttt{EVEREST} 2.0 to extract the K2 photometry.  For a detailed description of \texttt{EVEREST} see \citet{2016AJ....152..100L}.  Briefly, \texttt{EVEREST} employs an augmented PLD code to remove systematic errors introduced by the variable pointing during each K2 campaign.  It optimizes the pixel basis vectors using a principal component analysis.  This produces photometric precision comparable to that of the original Kepler mission. \cite{2016AJ....152..100L} found for stars brighter than Kepler magnitude 13, \texttt{EVEREST} produced a median precision with K2 data that was within a factor of 2 of the original Kepler mission.  \cite{2016AJ....152..100L} showed that \texttt{EVEREST} light curves have $\sim20\%$ less scatter on average compared to K2SFF\citep{K2SFF}.  They also show that \texttt{EVEREST} light curves have higher precision than those of K2SC \citep{K2SC} by $\sim10\%$ for bright stars and $\sim5\%$ for stars in the range $16 < Kp < 11$.  Overall \cite{2016AJ....152..100L} show that \texttt{EVEREST} outperforms other mainstream pipelines in scatter reduction.

In practice, \texttt{EVEREST} downloads the pixel level files using an appropriate aperture matched to the pixels that contain the majority of the stellar flux.  Average values from outside this aperture are treated as background and subtracted.  \texttt{EVEREST} splits the light curve into five even sections and removes significant outliers from each section.  Next, \texttt{EVEREST} determines what number of components is appropriate to use in the principal component analysis.  The order of the PLD to use for fitting as well as the number of divisions to use to divide the light curve are also determined.  Then these parameters are used to detrend the pixel level noise in the light curve. Figure \ref{fig:everest} shows an \texttt{EVEREST} detrended light curve for K2-167b.

\section{Analysis} \label{sec:analysis}

\subsection{Fitting K2 Transits}

We here illustrate our K2 fitting process, used on all four systems, using the example of K2-79b.  

The \texttt{EVEREST} pipeline was used to remove instrumental effects from K2 observations.  The standard deviation of the out-of-transit portion of the light curve was adopted as the photometric error per point.  Outliers, in the in-transit and out-of-transit regions, were removed using a threshold equal to 1.5 times the inter-quartile range of the region.  \texttt{BATMAN} was used to model the transits \citep{2015ascl.soft10002K}.  We fit the parameters for the time of first transit T0, period $Per$, and impact parameter $b$ in an initial fit using \texttt{emcee}, and MCMC python package \citep{2013PASP..125..306F}.  We apply Gaussian priors to the period, $T_{0}$, b.  For the period we used a central value and uncertainty from \cite{crossfield} and \cite{mayo}.  Generally, the T0 starting value and error approximation is estimated by eye.  The impact parameter $b$ uses a starting value of 0 and an uncertainty of 0.1.  In the case of K2-212 we use Gaussian priors for impact parameter based on \cite{2018Livingston}.  We then phase-fold the entire light curve based on these parameters.

\begin{figure}
\begin{center}
    \includegraphics[width=10cm]{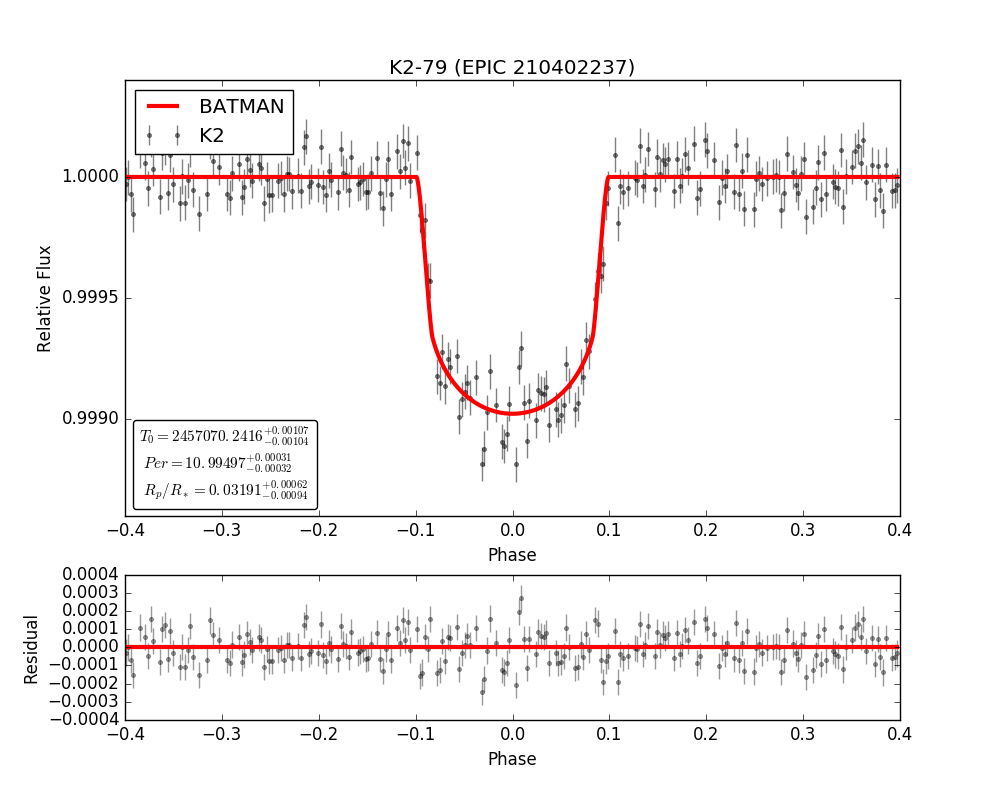}
        \caption{Result of MCMC fitting of a \texttt{BATMAN} model to K2 observations of K2-79b, and the residuals.}
    \label{fig:79bK2fit}
\end{center}
\end{figure}

\begin{figure}
    \includegraphics[width=8cm]{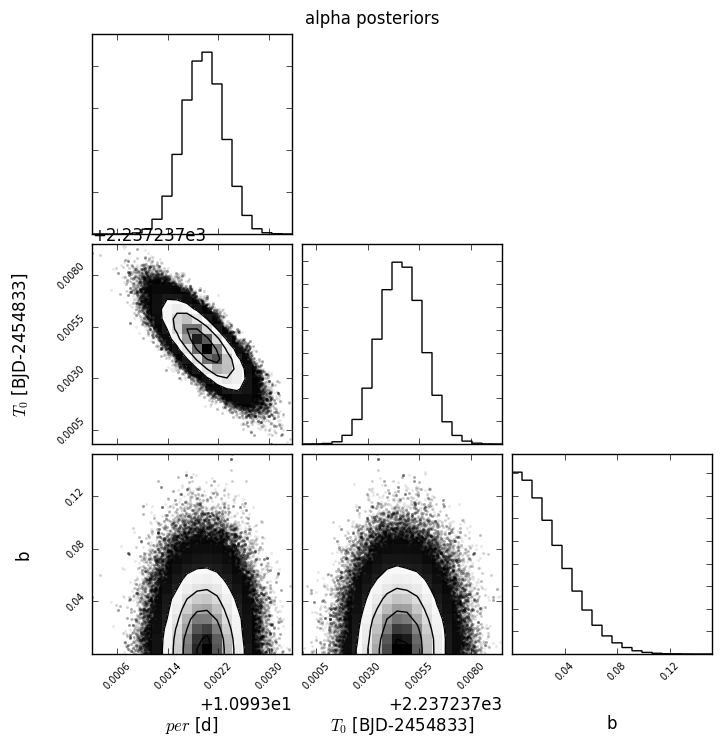}
    \caption{ The corner plot for the first set of parameters needed to fold the transit for K2-79b.}
    \label{fig:79bK2walka}
\end{figure}

\begin{figure}[ht]
    \includegraphics[width=8cm]{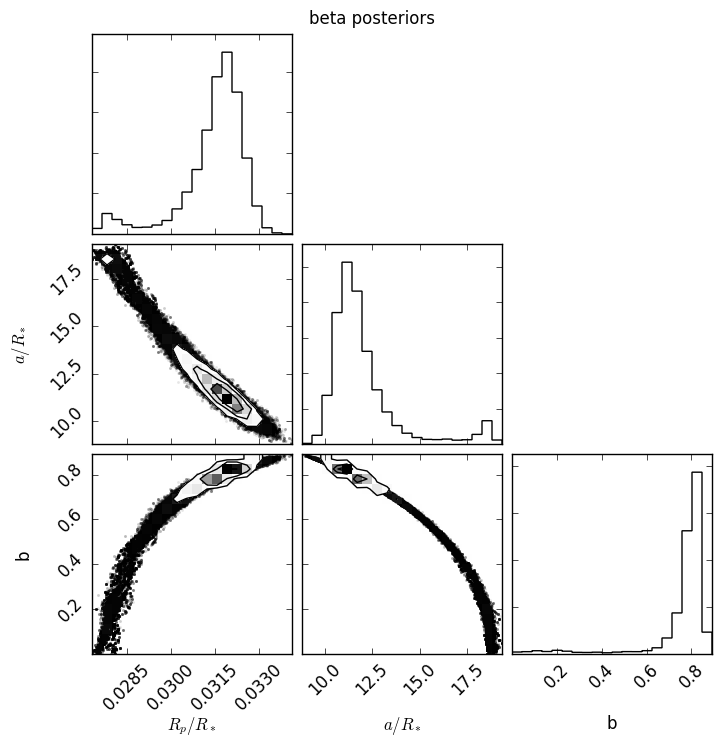}
    \caption{The corner plot for the second set of parameters fit after the light curve is phase-folded for K2-79b.}
    \label{fig:79bK2walkb}
\end{figure}

After the light curve was phase-folded, the MCMC procedure fit the time of transit center $T_0$, period $Per$, planet radius $R_p/R_*$, semi-major axis $a/R_*$, and impact parameter $b$.  Gaussian priors are applied to $R_p/R_*$ and $a/R_{*}$ using the starting value and uncertainties in \cite{crossfield} and \cite{mayo}.  Quadratic limb darkening coefficients were taken from \citet{2012AA...546A..14C}.   Figure~\ref{fig:79bK2fit} shows the phase folded fitted transit for K2-79b and the residuals after the best fit is subtracted from the data. The goodness of the fit can be visually assessed by the red line indicating the \texttt{BATMAN} model and the residuals in Figure \ref{fig:79bK2fit}. The residuals do not show any correlated structure, indicating a good fit. Figures \ref{fig:79bK2walka} and \ref{fig:79bK2walkb} show the resulting corner plot. The phase folding allows for more data points to define ingress and egress, aiding in the fit of $r_{p}/R_{*}$ and $a/R_{*}$.

\subsection{Fitting TESS Transits for K2-167b}

\begin{figure}[ht]
    \includegraphics[width=9cm]{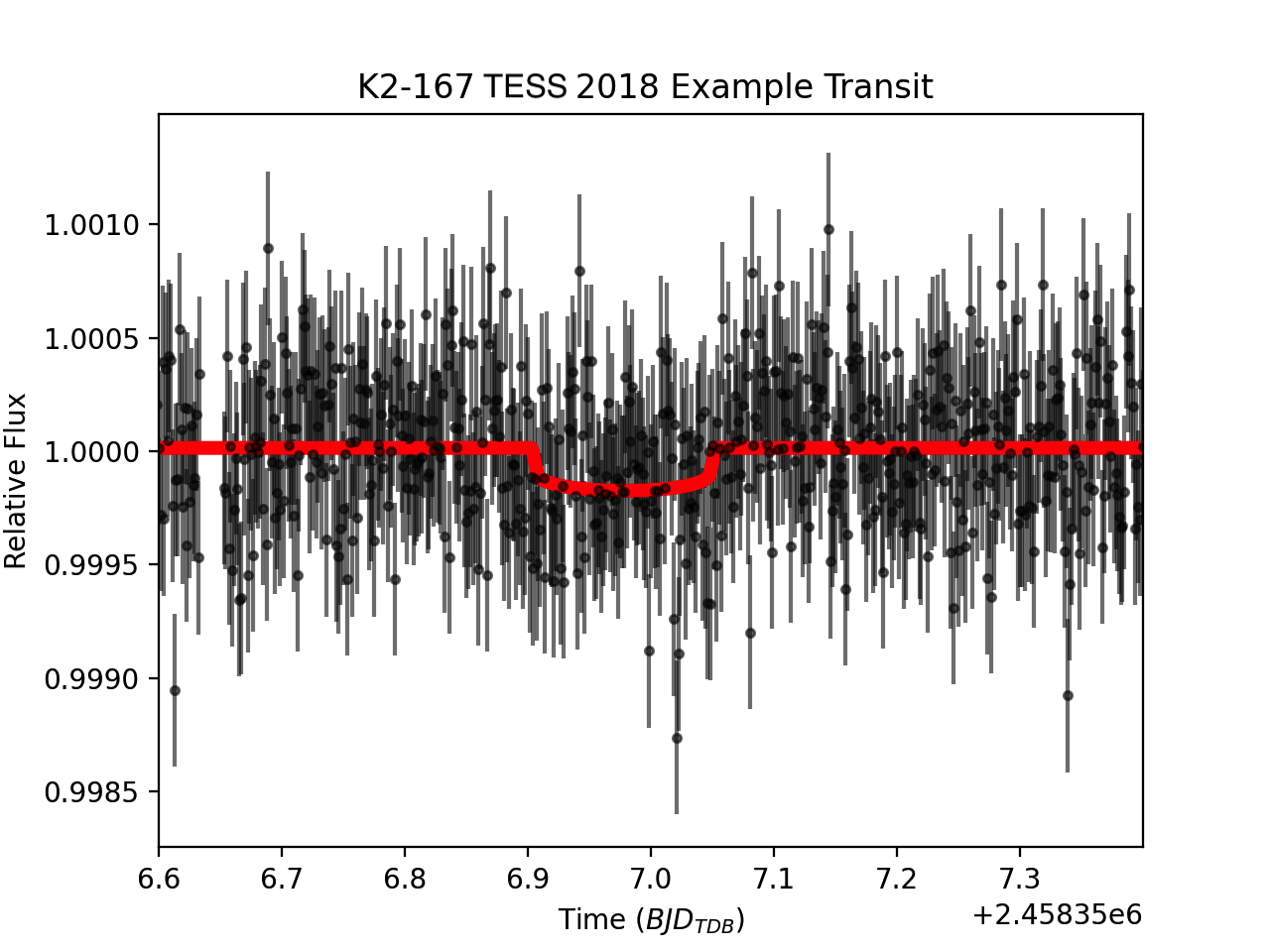}
        \caption{We present an example TESS transit over-plotted with the model in red used to determine the center of its transit time. The black points represent the PDCSAP fluxes and their associated uncertainties. }
    \label{fig:TESS_fit}
\end{figure}

TESS (Transiting Exoplanet Survey Satellite) observations were available only for K2-167.  We used the PDCSAP fluxes for that planet \citep{2015JATIS...1a4003R}. We then normalized any instrumental or stellar  variability using \texttt{Wotan}, a python based time series detrending package \citep{wotan}.  TESS obtained three transit light curves during the first 27-day visibility period for this planet in 2018.  Another transit was observed in 2020.  TESS has a much smaller collecting aperture than K2, so TESS's photon-limited precision for K2-167 is reduced compared to the K2 data.  Accordingly, we did not attempt to fit for the parameters that affect the shape of the K2-167 transit in the TESS data.  Instead, we adopt the published orbital parameters, and we fit for only the time of transit center. Figure~\ref{fig:TESS_fit} shows an example of the TESS photometery and fitted model for one transit.

\subsection{Fitting {\it Spitzer} Transits}

We calculated transit light curves for each planet using the \texttt{BATMAN} package \citep{2015ascl.soft10002K}.  We assumed circular orbits and used the values described in \citet{crossfield} as priors on the periods for planets K2-36c and K2-79b.  For K2-212b and K2-167b we set priors based on values from \citet{mayo}.  In each fit, we allowed $T_0$, $R_p/R_*$,$a/R_*$, and $b$ to vary using Gaussian priors.  We increased the uncertainties of \citet{mayo} and \citet{crossfield} by a factor of 3 to account for additional scatter in the Spitzer data. We applied uniform priors to b and $a/R_*$. We used a starting b value of 0.1 and the K2 value as a starting value for $a/R_{*}$.  We assumed fixed quadratic limb darkening coefficients for both {\it Spitzer} and K2, from \citet{2012AA...546A..14C} and \citet{2013AA...552A..16C}. Those coefficients were determined based on stellar parameters from \citet{mayo} and \citet{crossfield}.

The {\it Spitzer} data for each transit were recorded at a much faster cadence (0.4~seconds) than is necessary for fitting the transits.  We elected to bin the {\it Spitzer} data to 100 points; this makes the fitting code run much faster, and it also makes the shape of the transit visible by eye in the scatter of the data. The binning process will not affect the solution for the best-fit, providing that the binning is not sufficiently extensive to broaden the transit shape, and provided that we treat the error bars correctly.  We validated our binning procedure by also fitting data with only 2 points per bin, obtaining the same best-fit solution, albeit visually much less clear. The per-point precision of the {\it Spitzer} photometry was available from the {\it Spitzer} photometry code, and is based on the scatter in the data at the observed resolution after removal of the intra-pixel effect.  For the binned {\it Spitzer} observations, the error bar per binned point was calculated as the per-point precision divided by the square root of the number of points in each bin. This value is reported in Table \ref{tab:JointAnalysis_results}.

\begin{figure}[ht]
    \includegraphics[width=9cm]{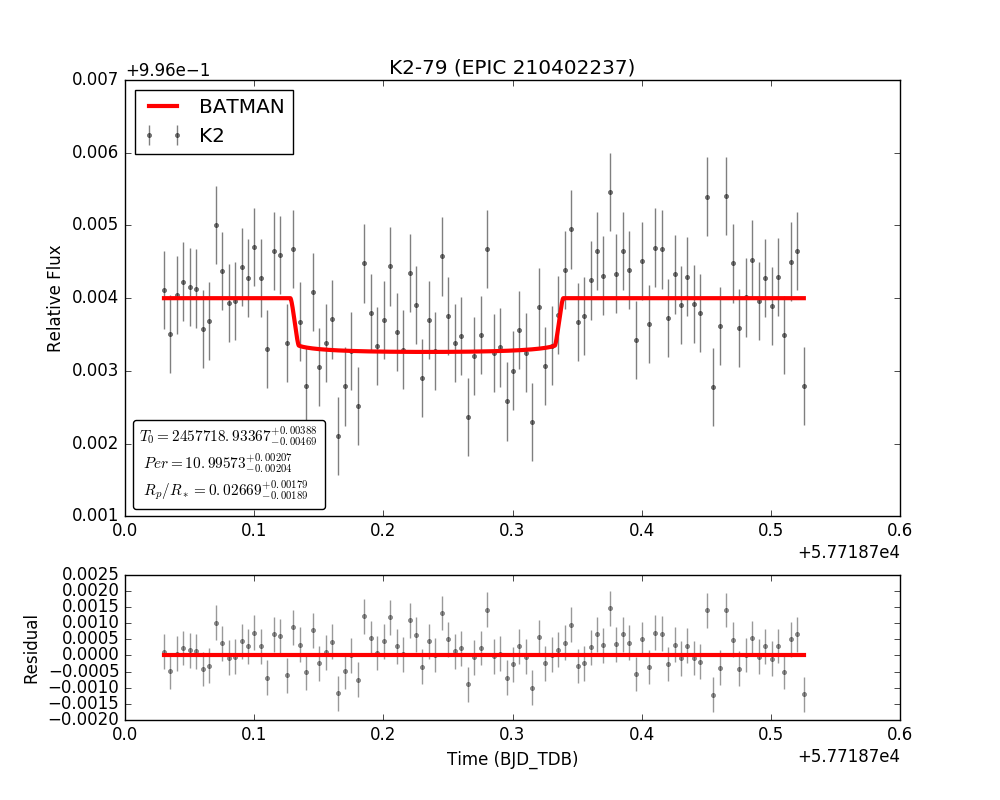}
    \includegraphics[width=9cm]{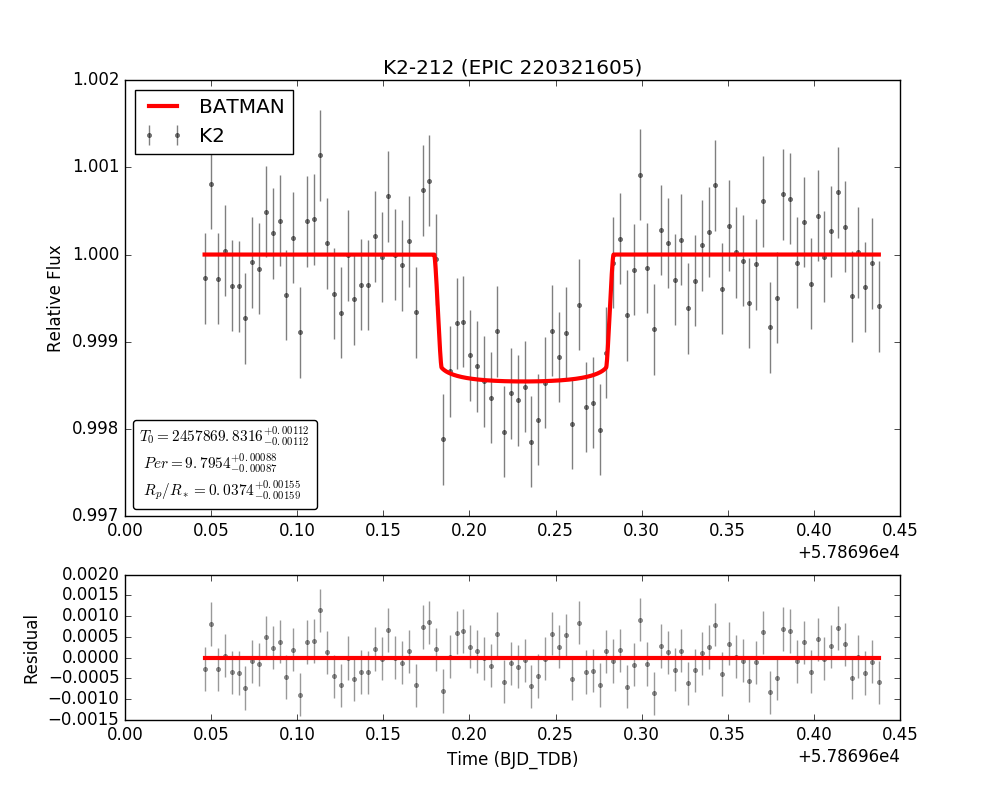}
    \caption{Results of the fit of a {\it Spitzer} transit of K2-79b in the upper plot and K2-212b in the lower plot.  The upper panels show the best fit to the photometry, and the lower panels show the residuals after the best fit is removed from the data.}
    \label{fig:79bSpit}
\end{figure}

We implement the \texttt{emcee} MCMC method to fit each transit model to the {\it Spitzer} light curves. The MCMC was run for 50000 steps.  A burn in of 2500 steps was used so the walkers could find the proper region of the parameter space to sample before it influenced the fit.   Figure~\ref{fig:79bSpit} shows the fit to the {\it Spitzer} data for the planets with the lowest (K2-79b) and highest (K2-212b) signal-to-noise ratio on the transit. In both cases, the residual plots in the lower panel show a flat line with points consistent with zero, indicating a good fit.

\begin{figure}[ht]
    \includegraphics[width=9cm]{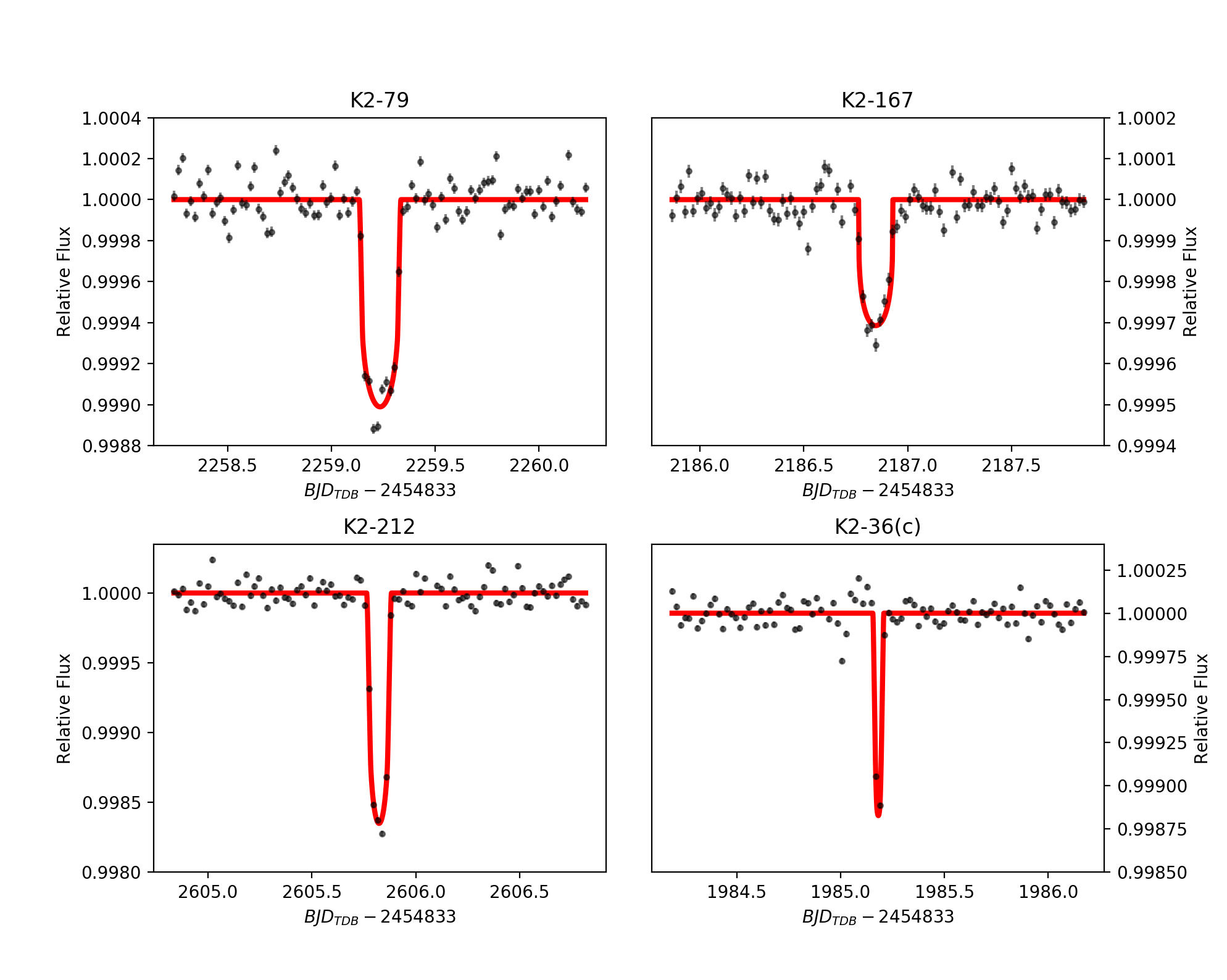}
    \caption{Selected individual K2 transits for all four planets, fitting the time of central transit using an MCMC and \texttt{BATMAN} model.}
    \label{fig:4Panel}
\end{figure}

\subsection{Joint Light Curve Analysis}\label{subsec:joint}

The joint analysis takes the best fit parameters from the independent K2 and {\it Spitzer} model fitting to use as Gaussian priors. We jointly fit $R_p/R_*$, per, $a/R_*$, and b. We apply Gaussian priors based on the results of both the K2 and Spitzer fits. This allows both data sets to jointly impact the priors of these parameters. The {\it Spitzer} data, having already been corrected with pixel level decorrelation, are binned using 100 bins and the K2 folded light curve corrected by \texttt{EVEREST} is incorporated into the fit. Because it is impractical to link the Spitzer PLD code with the K2 fitting procedure, our joint fit uses {\it Spitzer} error estimates that are based only on the scatter after the preliminary {\it Spitzer} fits, not including the uncertainties in removing {\it Spitzer's} intra-pixel effect.  Fortunately, the photon noise in the {\it Spitzer} photometry is dominant for these systems, so omission of instrumental uncertainties is acceptable.  We verified that assertion by calculating the covariance (or lack thereof) of each PLD pixel coefficient with the Spitzer transit depths.  From those covariances, we calculated the total effect of the pixel coefficients on the transit depths.  That total effect was largest ($4\%$ of the depth) for K2-79, and less than $1\%$ for the other three planets. Those effects are sufficiently small to guarantee that the photon noise dominates the precision of the transit depths in the joint fit.

Then, one likelihood function responds to variations in the parameters for two models, one model for the K2 observations and one model for the {\it Spitzer} transit, at the same time. Once the MCMC has run for 50,000 steps, the best fit parameters for both the {\it Spitzer} observations and the K2 observations are reported.  Since only K2-167b is available in the TESS data, we do not include TESS in the joint fits.  However, we use the TESS transits for K2-167b in the final stage of improving the orbital ephemeris.

\subsection{Testing for Mixing and Convergence}

To ensure accurate parameters from the fitting process, it is important to check that the Markov chains are well-mixed and converged.  We followed the general principles discussed by \citet{hogg2018}.  For each walker, we first visually inspected the sequence of samples for obvious problems. We then calculated the autocorrelation time of each walker using the IDL function $"A_{CORRELATE}"$, and we checked convergence of each walker using the diagnostic advocated by \citet{gewenke1992}.  

Because of the Markov property of the walkers, samples are intrinsically correlated from step-to-step.  We therefore expect the autocorrelation time to be significantly longer than a single sample, but much shorter than the length of the chain.  For our joint fits, we found that the autocorrelation properties of the chains were similar for all four planets, and all fitted parameters.  Specifically, the time for the autocorrelation function to drop by $1/e$ was typically about 4 samples, and always less than 10 samples.  That is short compared to the chain length of $5.0 \times 10^{4}$ samples, ensuring that the chains can adequately sample the parameter space. We also calculated the Geweke Z-score of each individual walker, that compares the average value in the first 10-percent of the chains to the last 50-percent of the chain.  The Z-score should be less than approximately 2, expressing that the average values differ by less than twice the standard deviation of their difference.  Our walkers had average Z-scores of less than 0.4 for all four planets in the orbital period, and less than 1.3 and 1.1 for the $T_0$ and planetary radius values, respectively.  Although only lack of convergence can be proven in principle \citep{hogg2018}, our Geweke Z-score values are all consistent with converged walkers. 

Although the autocorrelation times and Z-scores discussed above are for the joint (K2+Spitzer) fits, we also calculated the values for all walkers when fitting to the K2 and Spitzer data alone.  The results in those cases also led us to conclude that the chains were adequately mixed and converged.

\section{Results of the Joint Fits} \label{sec:results}

The result of the joint analysis is shown in Figure \ref{fig:JointFit1} located in the Appendix.  We also present an example corner plot for K2-79b in Figure \ref{fig:JointFitCorner} to show healthy posteriors and adequate sampling.  The same planet parameters are used in each of the red lines showing the planet model as seen by each telescope.  The only difference in shape comes from the slight difference in limb darkening as seen by K2 and {\it Spitzer} respectively.  The final planet parameters produced by the joint fit as shown in Table \ref{tab:JointAnalysis_results} in the Appendix \ref{app:figs}. We multiplied the ratio of planet-to-star radii from our fitting process, times the radii of the host stars from \citet{hardegree2020} to produce updated estimates of the physical radii of the planets characterized here. 

\subsection{Improving the Precision of Orbital Periods}

We improved the precision of the orbital period for each planet by leveraging the long time baseline provided by adjoining the {\it Spitzer} and TESS observations to K2.  In principle, the joint fitting process described above extracts an improved orbital period.  However, in practice we used an additional refinement to derive the most precise orbital periods.  For each individual transit, we used the result of the joint fit parameters to inform the \texttt{BATMAN} model, and only allowed the central time of transit and the radius of the planet to be re-fit by the MCMC. Figure \ref{fig:4Panel} shows examples of one individual transit for each planet. Then after the central time was found for each individual transit, it was plotted versus transit number as shown in Figure~\ref{fig:lin4Panel} in Appendix \ref{app:figs}. The latest point represents the central time of the {\it Spitzer} data (except for K2-167b where the last four points are from TESS).  The slope of the best-fit line represents the orbital period, obtained by conventional linear least-squares. Since the {\it Spitzer} data for each planet were taken at a much later time than K2, {\it Spitzer} and (for K2-167b, also TESS) helped to greatly increase the precision of the orbital period compared to using K2 data alone (see Sec.~\ref{sec:discussion}).  

Although the \texttt{EVEREST} photometry does an excellent job of removing noise due to pointing fluctuations, we did find that a few individual transits were outliers in the solutions for orbital period, likely because of extra noise.  We thereby zero-weighted those transits when calculating revised periods.  We dropped two K2 transits for K2-36c, none for K2-79b, one for K2-212b, and one for K2-167b. Table \ref{tab:analysis_results} delineates the central times used in calculating the refined orbital periods. Figure \ref{fig:lin4Panel} shows the timing residuals of the remaining transits after subtracting the least-squares line. The point shown after the break is the observation(s) by {\it Spitzer}/TESS.  

In most cases the scatter in the transit times on Figure~\ref{fig:lin4Panel} exceeds the error bars inferred from fitting the individual transit light curves.  That behavior could be caused by transiting timing variations, but spatial intensity fluctuations on the stellar disks could also play a role.  In any case, that scatter is accounted for in the errors that we derive for orbital periods.

\section{Simulating False-Positives}  \label{sec:vetting}

 \citet{crossfield} and \citet{mayo} calculated false-positive probabilities (FPP) for these systems, using \texttt{vespa} \citep{morton2015}. For K2-36c, K2-79b, K2-167b, and K2-212b, they find FPP values of $0$, $1.23 \times 10^{-9}$, $8.6 \times 10^{-4}$, and $2.0 \times 10^{-8}$, respectively.  Therefore these systems are highly likely to be real planets.  Nevertheless, it is interesting to determine to what extent the achromatic nature of the transits (K2 versus Spitzer) enables discrimination against false positives due to blended eclipsing binaries.  For each planet, we constructed 10,000 synthetic false positives, making them from the galactic population of stars in the direction of each planet.  We used the Trilegal model of stellar populations \citep{girardi}, synthesizing the stars in a 0.1-square degree field centered on the position of each exoplanet. 

We construct synthetic BEBs by first estimating that the primary transit of a putative BEB has a relative depth of 0.5 when not blended with the exoplanet host star. We use that depth to determine a K2 magnitude for primary stars in the BEB that would mimic the planetary transit when they are diluted by the exoplanetary host star.  We choose BEB primary stars within $\pm$0.3-magnitudes of that value.  For a randomly selected primary star in that sample, we make a BEB by adopting a uniform distribution of secondary masses (see \citealp{greklek} for a discussion of that choice), and we choose the secondary star randomly from a uniform distribution of secondary masses less than the primary star mass in the Trilegal sample.  We then calculate the real unblended primary transit depth of that BEB based on Phoenix model atmospheric fluxes \citep{husser} of the primary and secondary star in the K2 and Spitzer bands, integrating over the bandpass response functions.  For each binary, we then (in effect) change it’s distance slightly so that the blended K2 transit depth exactly matches the TESS planetary depth.  Those adjustments are sufficiently small so that the results remain statistically consistent with the nature of the Trilegal stellar population.

For each synthetic BEB, we calculate the false-positive transit depth in the Spitzer bandpass as well as the K2 bandpass.  We count the fraction of cases when the absolute value of the difference in those synthetic depths exceeds the uncertainty in the observed transit depth difference (K2 minus Spitzer) by more than 3 times the observed standard deviation of the difference.  We find that, for all of the exoplanets we analyze, less than 1\% of the synthetic false-positives could be discriminated on that basis. Therefore our Spitzer observations, while valuable as adding statistical weight to the observed exoplanetary radii and ephemerides, cannot improve the FPP values determined by \citet{crossfield} and \citet{mayo}.

\section{Summary and Discussion} \label{sec:discussion}

In this analysis, we were able to consistently fit for improved orbital periods, and radii, using data from both K2 and Spitzer.  In the case of K2-167b we were also able to incorporate four transits from TESS.  In principle, consistently fitting radii between K2 and {\it Spitzer} could demonstrate that the transits are achromatic, as expected for exoplanets \citep{2015ApJ...804...59D}, as opposed to false-positives from blended stellar eclipsing binaries. Investigating that aspect, we found that the signal-to-noise for these shallow transits is not sufficient to make that discrimination. 

Assuming that these are real planets (as per \citealp{crossfield} and \citealp{mayo}), we were able to increase the precision of their radii and orbital periods, as we now discuss.

Our new orbital periods are within the error envelope of previous measurements \citep{crossfield, damasso, mayo}, the difference being less than twice the previous uncertainty (i.e., $< 2\sigma$) in all cases.  However, our new values are significantly more precise than the above investigations were able to determine. Our largest improvement in precision of the orbital period is for K2-79b, where our period uncertainty is a factor of 28 less than given by \citet{crossfield}.  For K2-212b, our period uncertainty is more than a factor of 10 less than reported by \citet{mayo}.  

In the case of K2-167b we did not have the full {\it Spitzer} transit. The complete egress is missing due to the uncertainty in scheduling the {\it Spitzer} observations. Fortunately, our fitting process was still able to estimate the time of inferior conjunction $T_0$, and we also used the TESS transits for that planet. Recently, an updated ephemeris for K2-167b was derived by \citet{ukwa} who incorporate some TESS transits, but not the {\it Spitzer} transit or the most recent TESS transits.  Our orbital period for K2-167b agrees with the value from \citet{ukwa} within $2\sigma$, and both values are more than an order of magnitude more precise than the original estimate from \citet{mayo}.  

Our smallest improvement in period precision is a factor of 3 for K2-36c, comparing to \citet{damasso}.  Those authors used RV data that evidently were useful to increase their precision for the orbital period without using the {\it Spitzer} transit.

 Our joint-fit radii for these planets are listed in Table~ \ref{tab:JointAnalysis_results}. In all cases our values are within the uncertainty envelope of previous measurements. We find K2-36c and K2-167b to be within $0.7\sigma$ and $1.0\sigma$ of \citet{sinukoff} and \citet{mayo}, respectively. K2-79b is a $0.9\sigma$ revision of the radius given by \citet{crossfield}, and our result for K2-212b revises the value from \citet{mayo} upward by $1.4\sigma$. In all cases, we are confident that our new radii are improvements. 

\section{Implications for Future Work} \label{sec:conclusion}

Our results have implications for potential atmospheric characterization of these planets using JWST. We estimated masses, and calculated the transmission spectroscopy metric (TSM, \citealp{kempton}) of these planets. A common (though simplistic) scaling law is that a planet's mass scales with radius as Mp $\propto Rp^{2.06}$ \citep{lissauer}. Using masses from that relation, we obtain TSM = 43, 22, 27, and 35 for K2-36c, 79b, 212b, and 167b, respectively.  K2-36c is the best for atmospheric characterization, especially given that there is another planet in that system, potentially enabling comparative studies.  \citet{guo} found K2-36c to be the 10th most favorable "warm Neptune", based on the TSM.  Our updated results move it down to $>$\,20th, but still a very interesting case.  

We explored to what degree the TSM may change if other mass-radius relations apply, and we note that K2-36(c) has a measured mass from \citet{damasso}.   Using the mass-radius relation from \citet{chen2017}, we calculate the TSM for K2-36c to be 45, but that improves to TSM=57 using the measured mass (albeit that has a large error - $29$\%). \citet{kempton} predicted that TSM values of 90 or greater would be characteristic of the best sub-Neptunes, and indeed the best planets listed by \cite{guo} are consistent with that prediction. We conclude that K2-36c is not likely to be among the very best of the TESS sub-Neptunes for atmospheric characterization, for any plausible mass.  Nevertheless, it falls among the many sub-Neptunes wherein water vapor absorption could be detected in transit using the NIRISS instrument on JWST \citep{louie2018}.

\begin{figure}[ht]
    \includegraphics[width=9cm]{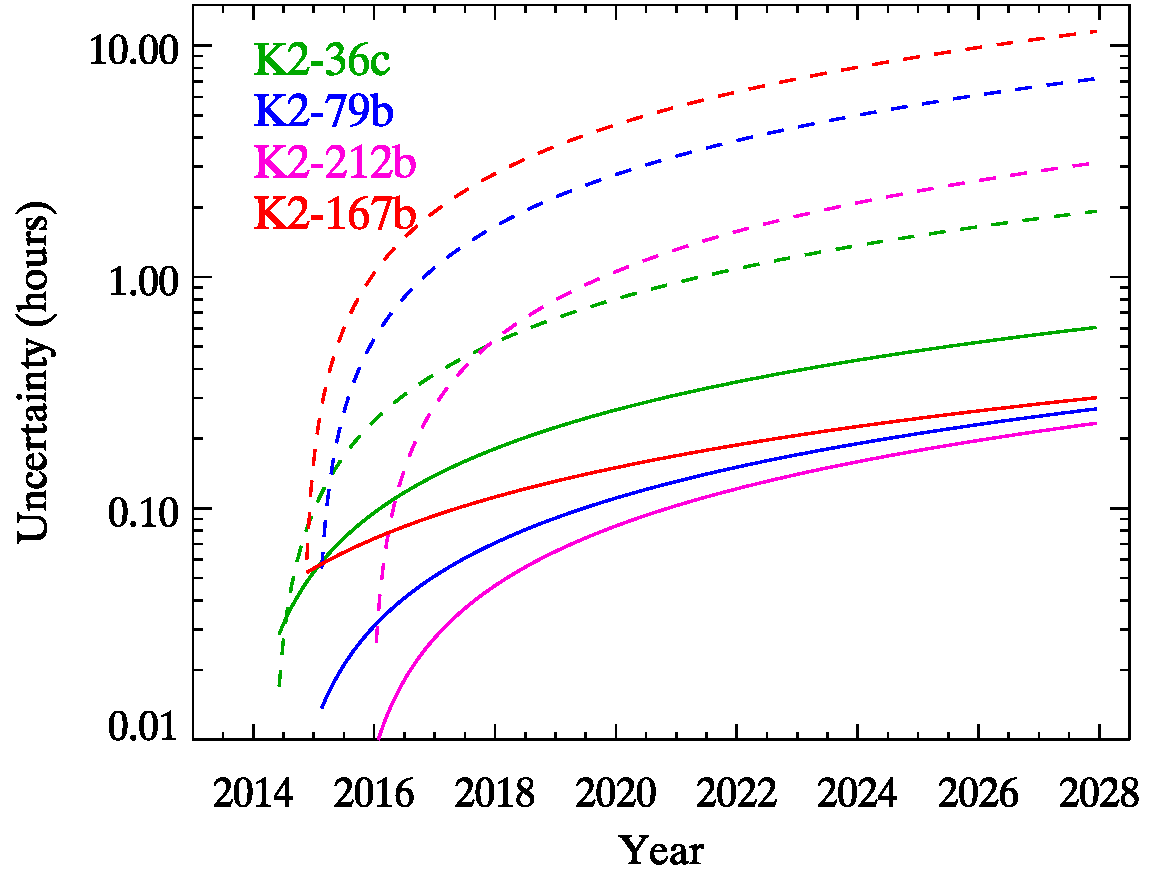}
    \caption{Uncertainty in transit time versus year, based on previous studies (dashed lines, \citealp{crossfield, damasso, mayo}), and this work (solid lines).  The transit time uncertainties include both the period as well as the error on the zero epoch transit ($T_0$).}
    \label{fig:precision}
\end{figure}

Our increased precision on the orbital periods of these planets will facilitate scheduling future transit observations.  For example, if these planets are observed more than a year after the JWST launch (e.g., 1/2023), the $1\sigma$ uncertainty in transit time will be 23, 10, 12, and 8 minutes for K2-36c, K2-79b, K2-167b, and K2-212b, respectively. For comparison the $1\sigma$ uncertainties based on \citet{crossfield}, \citet{damasso}, and \citet{mayo} would be 74, 271, 434, and 111 minutes, respectively, making the transits previously problematic to observe with JWST.  Our transit time precision versus future year of observation is shown in Figure~\ref{fig:precision}.

\acknowledgements
\section*{Acknowledgements}
This paper includes data collected by the Kepler mission and obtained from the MAST data archive at the Space Telescope Science Institute (STScI). Funding for the Kepler mission is provided by the NASA Science Mission Directorate. STScI is operated by the Association of Universities for Research in Astronomy, Inc., under NASA contract NAS 5–26555. This paper also includes data collected with the TESS mission, obtained from the MAST data archive at STScI. Funding for the TESS mission is provided by the NASA Explorer Program. This work is also based [in part] on observations made with the Spitzer Space Telescope, which was operated by the Jet Propulsion Laboratory, California Institute of Technology under a contract with NASA. Support for this work was provided by NASA through an award issued by JPL/Caltech.  

This research has made use of the VizieR catalogue access tool, CDS, Strasbourg, France. The original description of the VizieR service was published in A\&AS 143, 23. C.K.H. acknowledges support from the National Science Foundation Graduate Research Fellowship Program under Grant No. DGE1752814. The authors would also like to acknowledge the use of software packages \texttt{EVEREST}, \texttt{BATMAN}, \texttt{emcee}, \texttt{Numpy}, \texttt{Scipy}, and \texttt{Matplotlib}.  We thank Professor Scott Gaudi for helpful comments on a draft of this paper. We also thank the anonymous referee for through comments that strengthened this paper.

\bibliography{exopapers}

\software{matplotlib \citep{Hunter:2007}, Numpy \citep{harris2020array}, Scipy \citep{2020SciPy-NMeth}, emcee \citep{2013PASP..125..306F}, \texttt{EVEREST} \citep{2016AJ....152..100L}, BATMAN \citep{2015ascl.soft10002K}, Wotan \citep{wotan}}

\appendix
\section{Large Figures and Tables} \label{app:figs}

\begin{figure*}[ht]
	\centering
    \begin{minipage}[b]{0.49\textwidth}
    	\includegraphics[width=\textwidth]{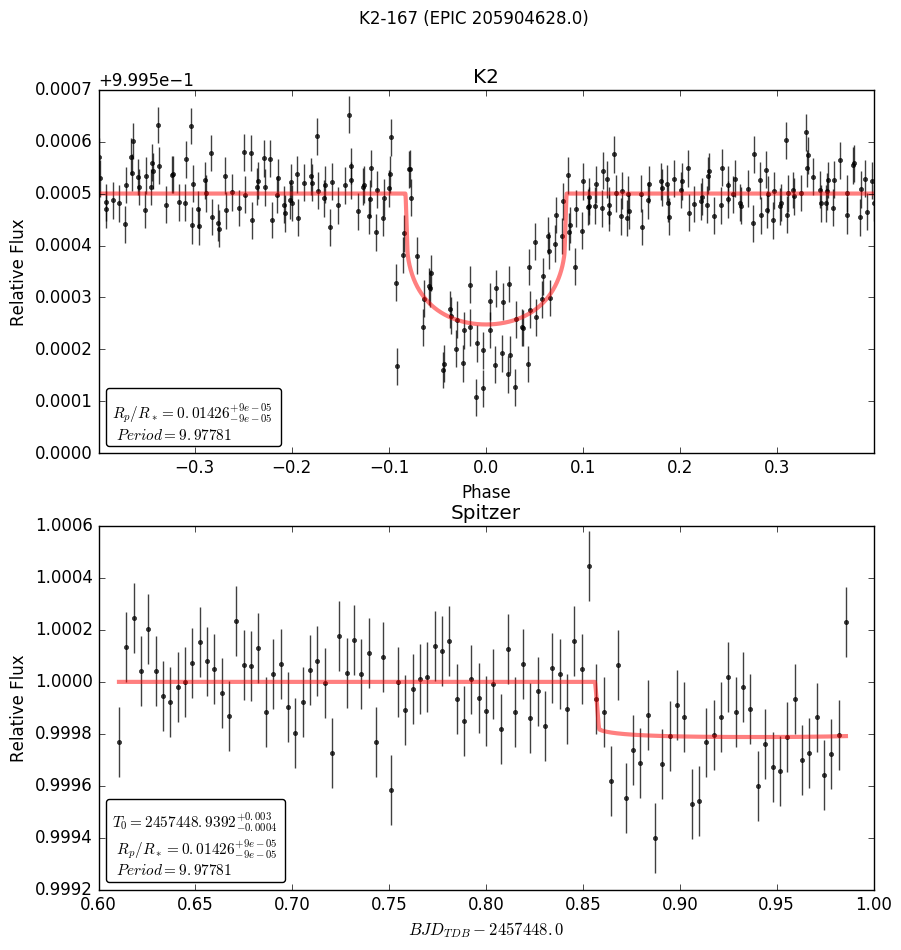}
    \end{minipage}
    \begin{minipage}[b]{0.49\textwidth}
    	\includegraphics[width=\textwidth]{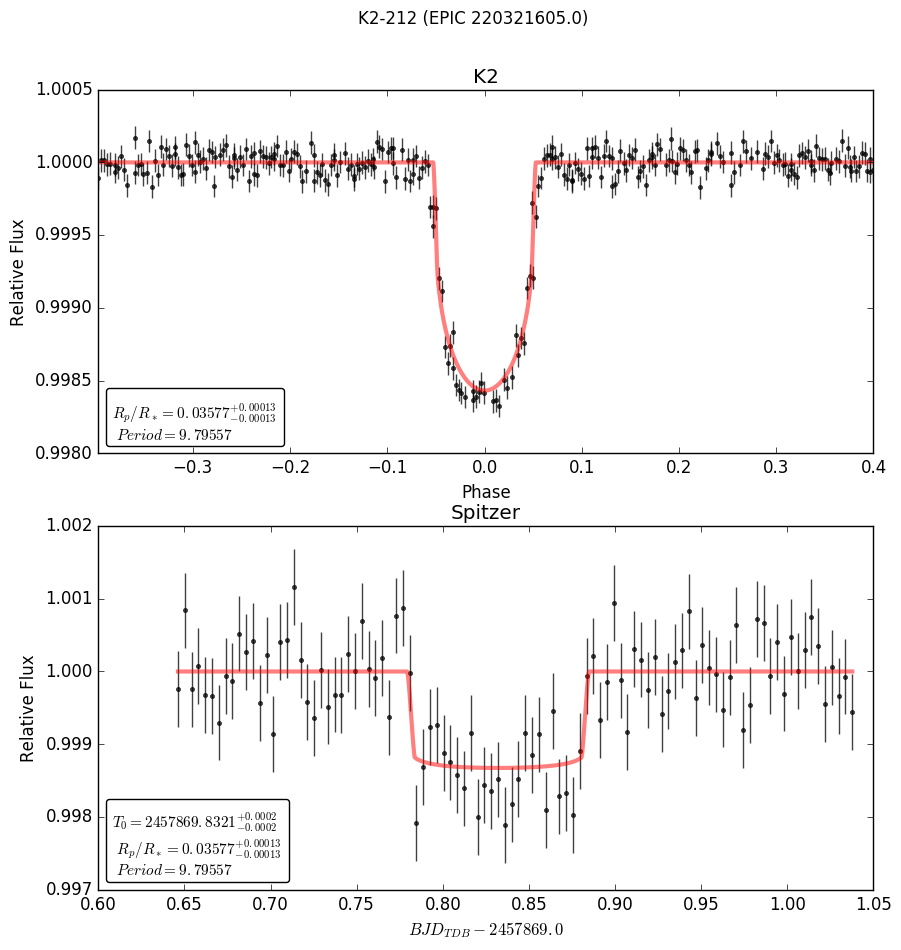}
    \end{minipage}
    \begin{minipage}[b]{0.49\textwidth}
    	\includegraphics[width=\textwidth]{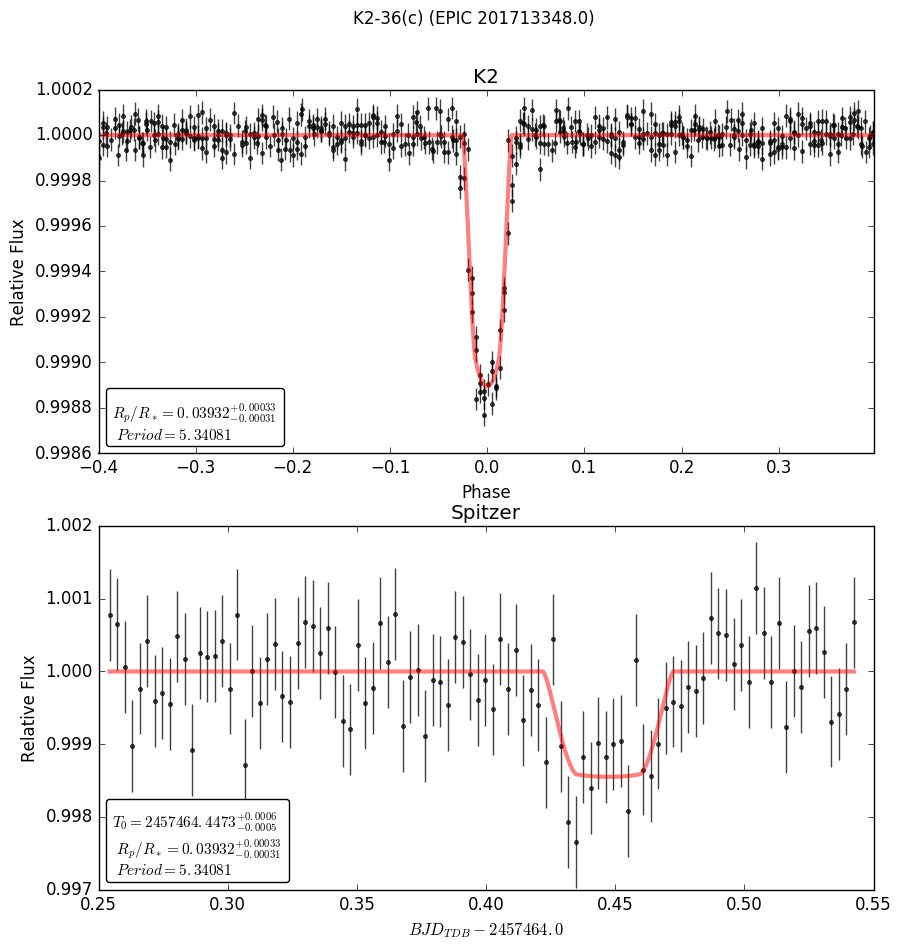}
    \end{minipage}
    \begin{minipage}[b]{0.49\textwidth}
    	\includegraphics[width=\textwidth]{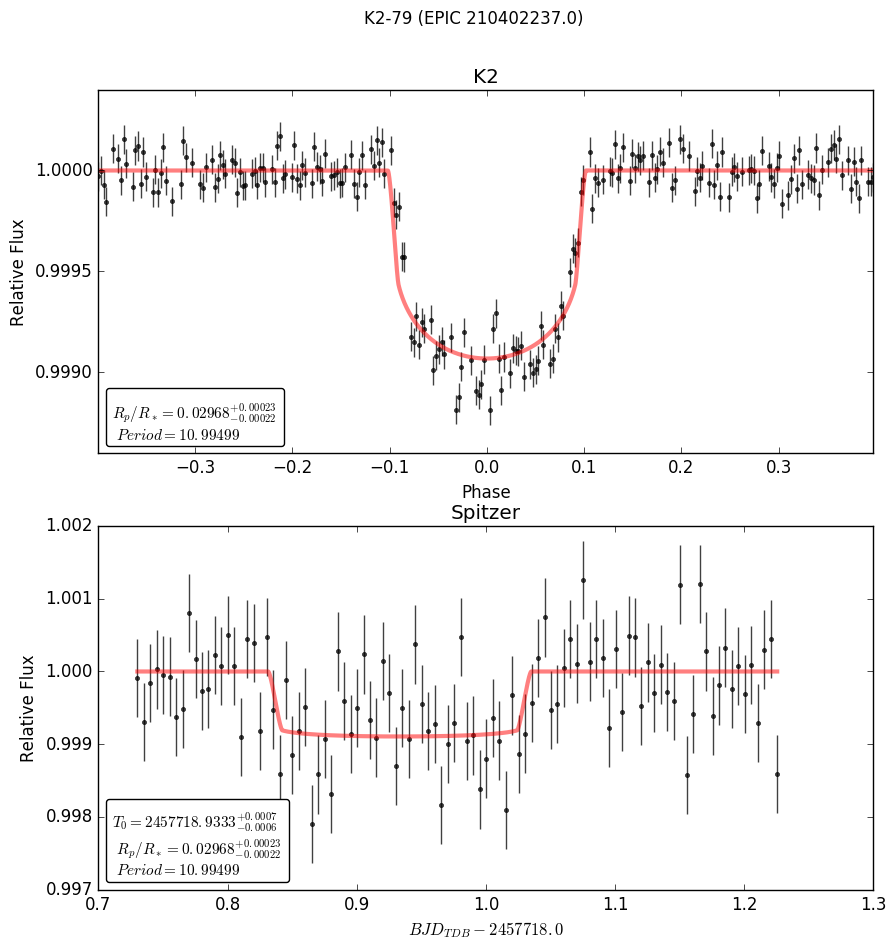}
    \end{minipage}
    
    \caption{Here the result of the joint analysis of the K2 and Spitzer observations are shown for our four targets. The joint model and the phase folded K2 data are shown on the upper panel. The joint model and the single Spitzer observation are displayed on the lower panel. The joint model is based on the same planet parameters in each panel, only the limb darkening is changed.}
   \label{fig:JointFit1}
\end{figure*}

\begin{figure*}[ht]
	\centering
    \begin{minipage}[b]{0.9\textwidth}
    	\includegraphics[width=\textwidth]{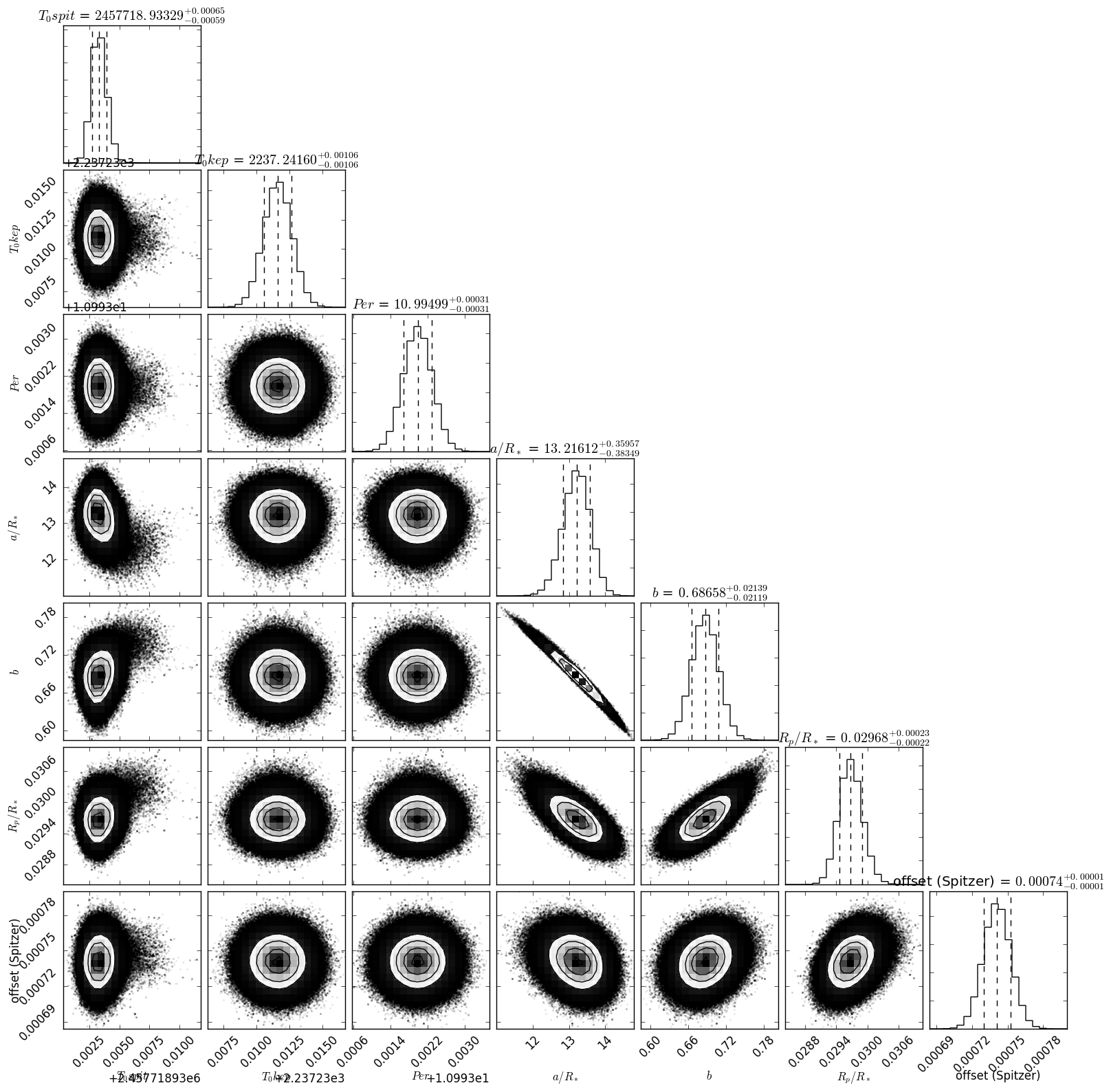}
    \end{minipage}
    
    \caption{Corner plot for the K2 and Spitzer joint fit for K2-79b.}
    \label{fig:JointFitCorner}
\end{figure*}

\begin{deluxetable*}{lcccc}[ht]
\tablecaption{Summary of Individual and Joint Fit Results. The planet radius is estimated using transit depth and the \cite{hardegree2020} stellar radii. \label{tab:JointAnalysis_results}}
\tablehead{
\colhead{} & \colhead{Spitzer} & \colhead{K2}
& \colhead{Joint Fit} & \colhead{Units}
}
\startdata
\multicolumn{5}{l}{\textbf{K2-36c}}	\\
$R_p/R_*$           	& $0.03396^{+0.003}_{-0.0029}$ & $0.0402^{+0.0006}_{-0.0006}$ & $0.0393^{+0.00033}_{-0.00031}$ & $-$ \\
$R_p$            		& $-$ & $-$ & $3.12^{+0.22}_{-0.21}$ & $R_{\oplus}$ \\
$a/R_*$					& $30.77^{+ 8.46}_{-10.094}$ & $15.12^{+0.73}_{-0.65}$ & $15.32^{+0.49}_{-0.57}$ & $-$ \\
$b$ 					& $0.44^{+ 0.35}_{-0.30}$ & $0.94037^{+0.005}_{-0.006}$ & $0.9345^{+0.0039}_{-0.0036}$ & $-$ \\
$T_0$					& $2457464.447^{+0.0027}_{-0.0035} $ & $2456812.8401^{+0.0005}_{-0.0007}$ & $2456812.83943 \pm 0.00119$ & BJD (TDB) \\
$Per$					& $5.340715^{+0.00033}_{-0.00033}$ & $5.3408154^{+8.0e-05}_{-6.16e-05}$ & $5.3410507 \pm 0.0000257$& Days \\
$\sigma_{flux}$ 		& $0.0006$ & $0.0000502$ & $-$ & $-$ \\
Transit Depth           & $ 0.0011 \pm 0.00023 $  & $0.00162 \pm 0.00005$ & $0.001545 \pm 0.000026$   & $-$ \\
\hline
\multicolumn{5}{l}{\textbf{K2-79b}}	\\
$R_p/R_*$ 			& $0.0267^{+0.0018}_{-0.0019}$ & $0.0319^{+0.00062}_{-0.00093}$ & $0.02968^{+0.00028}_{-0.00022}$ & $-$ \\
$R_p$				& $-$ & $-$ &$4.05^{+0.25}_{-0.23}$ & $R_{\oplus}$ \\
$a/R_*$					& $15.7^{+1.5}_{-3.5}$ & $11.5^{+1.3}_{-0.80}$ & $13.22^{+0.36}_{-0.38}$  & $-$ \\
$b$ 					& $0.39^{+0.31}_{-0.27}$	& $0.80^{+0.03}_{-0.06}$  & $0.68^{+0.02}_{-0.02}$ & $-$ \\
$T_0$					& $2457718.93367^{+0.0038}_{-0.0046}$ & $2457070.2416^{+0.001}_{-0.001}$ &  $2457070.24341 \pm 0.00057$ & BJD (TDB) \\
$Per$					& $10.9958^{+0.002}_{-0.002}$ & $10.99496^{+0.00032}_{-0.00034}$ &  $10.994748 \pm 0.000025$ & Days \\
$\sigma_{flux}$ 		& $0.00054$ & $0.00007$ & $-$ & $-$ \\
Transit Depth         & $ 0.00071 \pm 0.000096 $ & $0.00102 \pm 0.000039 $ & $0.00088 \pm 0.000016$    & $-$ \\
\hline
\multicolumn{5}{l}{\textbf{K2-212b}}	\\
$R_p/R_*$ 			& $0.0374^{+0.00155}_{-0.0015}$ & $0.0347^{+0.00018}_{-0.00018}$ & $0.03577^{+0.00013}_{-0.00013}$ & $-$ \\
$R_p$				& $-$ & $-$ & $2.65^{+0.20}_{-0.18}$ & $R_{\oplus}$ \\
$a/R_*$					& $29.3^{+1.8}_{-4.7}$ & $26.83^{+0.18}_{-0.18}$ & $27.97^{+0.134}_{-0.14}$ & $-$ \\
$b$ 					& $0.33^{+0.27}_{-0.23}$	& $0.37995^{+0.0099}_{-0.0098}$  & $0.42028^{+0.008}_{-0.008}$ & $-$ \\
$T_0$					& $2457869.8315^{+0.0011}_{-0.0011}$ & $2457399.6477^{+0.002}_{-0.0027}$ & $2457399.64109 \pm 0.00039$ & BJD (TDB) \\
$Per$					& $9.7953^{+0.00087}_{-0.00087}$ & $9.7956^{+0.0003}_{-0.00029}$ & $9.795635 \pm 0.000021$ & Days \\
$\sigma_{flux}$ 		& $0.00052$ & $0.00008$ & $-$ & $-$ \\
Transit Depth         & $ 0.0014 \pm 0.0001 $ & $0.0012060 \pm 0.000012$ & $0.0012797 \pm 0.000009$  & $-$ \\
\hline
\multicolumn{5}{l}{\textbf{K2-167b}}	\\
$R_p/R_*$ 			& $0.01545^{+0.00091}_{-0.00096}$ & $0.013985^{+0.00012}_{-0.000082}$ & $0.0142568^{+8.6e-05}_{-8.6e-5}$ & $-$ \\
$R_p$				& $-$ & $-$ & $2.34^{+0.14}_{-0.12}$ & $R_{\oplus}$ \\
$a/R_*$					& $23.6^{+2.17}_{-5.62}$ & $19.4000007^{-7.34e-05}_{-6.8e-05}$ & $19.400001^{+7.3e-05}_{-7.33e-5}$ & $-$ \\
$b$ 					& $0.39^{+0.32}_{-0.27}$	& $-0.000028^{+	0.0001}_{-8.5e-05}$ & $6.3e-05^{+7.3e-05}_{-4.4e-5}$ & $-$ \\
$T_0$					& $2457448.921^{+0.002}_{-0.002}$ & $2456979.9386^{+0.0006}_{-0.0006}$ & $2456979.9385  \pm 0.0022$ & BJD (TDB) \\
$Per$					& $9.977^{+0.0029}_{-0.0029}$ & $9.9778^{+0.0002}_{-0.0002}$ & $9.9785275 \pm 2.16e-5$ & Days \\
$\sigma_{flux}$ 		& $0.00013$ & $0.00003$ & $-$ & $-$ \\
Transit Depth           & $0.00024 \pm 0.00003$ & $0.0001956 \pm 0.000003$ & $ 0.000203 \pm 2.45 e-6 $    & $-$ \\
\enddata

\end{deluxetable*}

\begin{figure*}[ht]
	\centering
    \begin{minipage}[t]{\textwidth}
    	\includegraphics[width=\textwidth]{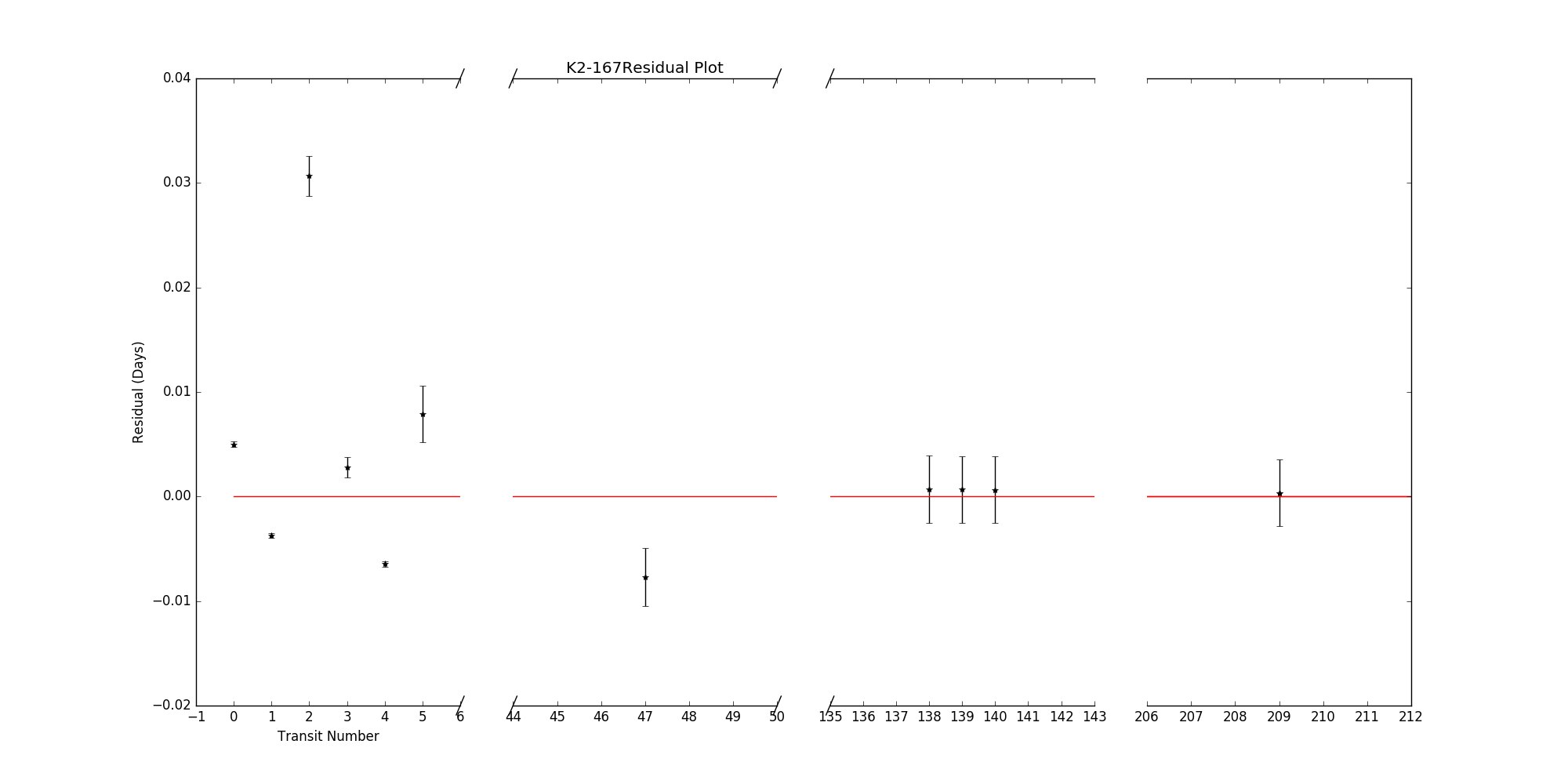}
    \end{minipage}
    \begin{minipage}[t]{0.49\textwidth}
    	\includegraphics[width=\textwidth]{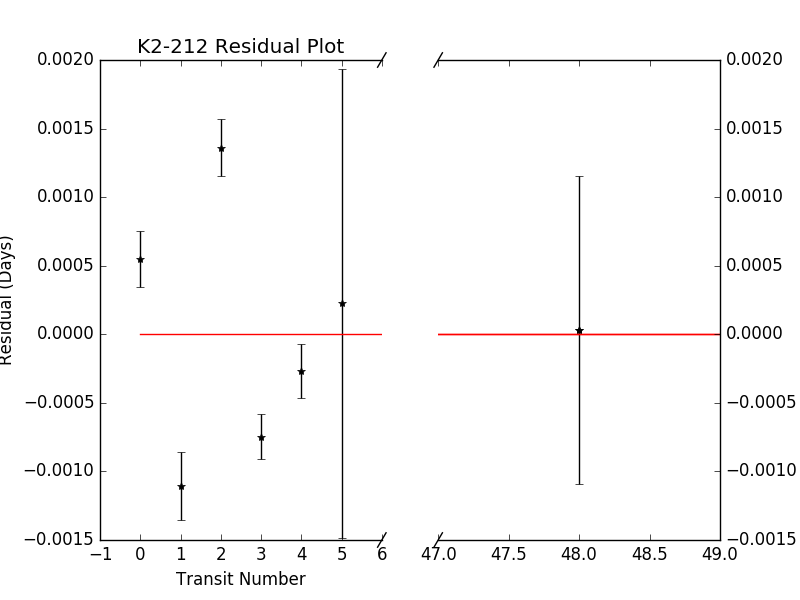}
    \end{minipage}

    \begin{minipage}[t]{0.49\textwidth}
    	\includegraphics[width=\textwidth]{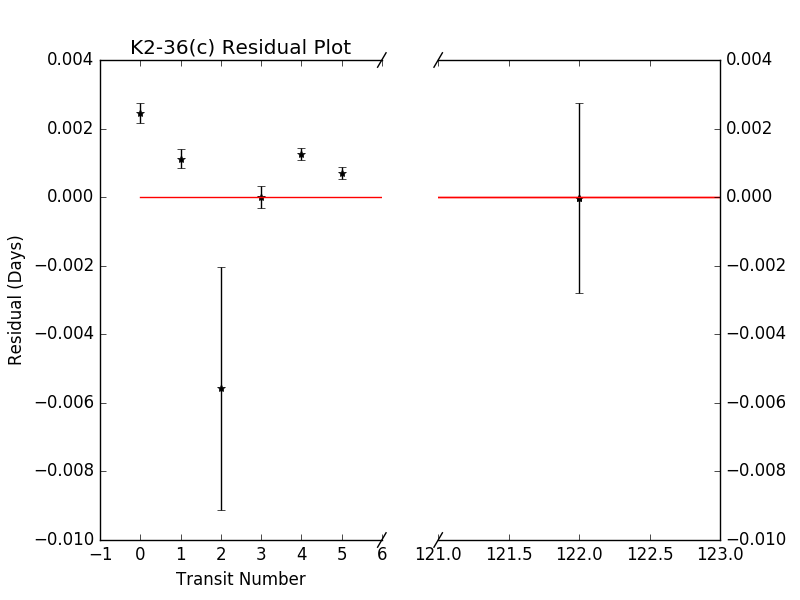}
    \end{minipage}
    \begin{minipage}[t]{0.49\textwidth}
    	\includegraphics[width=\textwidth]{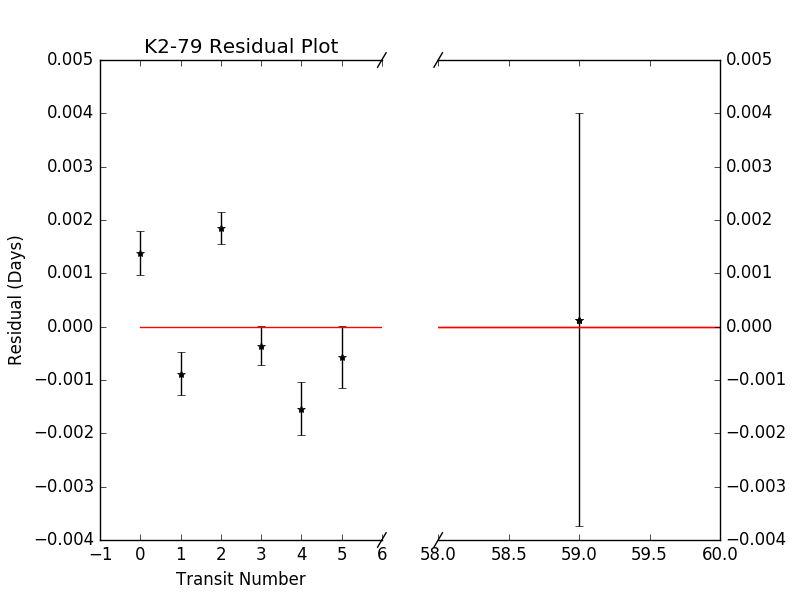}
    \end{minipage}
    \label{fig:JointFit2}
    \caption{These residual plots show how well each period represents the measured transit times for each system. The transits before the line breaks were observed by K2. The points after were observed by {\it Spitzer}. K2-167b also has two additional observation by TESS resulting in another 4 transits. These observations are shown after the right-most axis break. The central times used to create this figure are shown in table \ref{tab:analysis_results}}
    
    \label{fig:lin4Panel}
\end{figure*}

\begin{deluxetable*}{lcc}[ht]
\tablecaption{Summary of Analysis Results. Below are the reported transit centers and radii for each transit. Continued on next page. Several individual transits had large uncertainties in their central times although we were able to make transit depth determinations. Therefore the central times of a few transits has been removed.  Times are reported in $BJD_\textrm{TDB}$. \label{tab:analysis_results}}
\tablehead{
\colhead{Transit Number} & \colhead{Central time} & \colhead{Rp/R*}
}
\startdata
\textbf{\large{K2-36c}} & &	\\
\textbf{K2 Observations} & &	\\
Transit 0           	& $2456812.841922 \pm 0.000270$ & $0.041636 \pm 0.000272$ \\
Transit 1           	& $2456818.181645 \pm 0.000287$ & $0.042071 \pm 0.000266$ \\
Transit 2            	& $ - $ & $ - $  \\
Transit 3				& $2456828.862578 \pm 0.000323$ & $0.040846 \pm 0.000267$ \\
Transit 4 				& $2456834.204920 \pm 0.000162$ & $0.042142 \pm 0.000275$ \\
Transit 5				& $2456839.545393 \pm 0.000188$ & $0.042678 \pm 0.000303$ \\
Transit 6				& $2456844.886742 \pm 0.000281$ & $0.041624 \pm 0.000289$ \\
Transit 7				& $ - $ & $0.041653 \pm 0.444262$ \\
Transit 8				& $2456855.568375 \pm 0.000304$ & $0.043039 \pm 0.000264$ \\
Transit 9				& $2456860.905830 \pm 0.000379$ & $0.039550 \pm 0.000382$ \\
Transit 10				& $2456866.248332 \pm 0.000170$ & $0.042439 \pm 0.000292$ \\
Transit 11				& $2456871.589899 \pm 0.000261$ & $0.047025 \pm 0.000253$ \\
Transit 12				& $2456876.929997 \pm 0.000160$ & $0.042487 \pm 0.000276$ \\
Transit 13				& $2456882.265958 \pm 0.000169$ & $0.038974 \pm 0.000298$ \\
Transit 14				& $2456887.614063 \pm 0.000194$ & $0.041480 \pm 0.000246$ \\
\textbf{Spitzer Observation}& &	\\
Transit 122				& $2457464.447630\pm 0.002777$ & $0.033962 \pm 0.002885$ \\
\hline
\textbf{\large{K2-79b}}& &	\\
\textbf{K2 Observations}& &	\\
Transit 0           	& $2457070.244748 \pm 0.000399$ & $0.031000 \pm 0.000164$ \\
Transit 1           	& $2457081.237340 \pm 0.000398$ & $0.030910 \pm 0.000157$ \\
Transit 2            		& $2457092.234740 \pm 0.000371$ & $0.032399 \pm 0.000175$ \\
Transit 3					& $2457103.227306 \pm 0.000324 $ & $0.031977 \pm 0.000142$ \\
Transit 4 					& $2457114.220866 \pm 0.000466$ & $0.027713 \pm 0.000198$ \\
Transit 5					& $2457125.216603 \pm 0.000480$ & $0.030309 \pm 0.000162$ \\
\textbf{Spitzer Observation}& &	\\
Transit 59					& $2457718.933674 \pm 0.003877$ & $0.026686 \pm 0.001792$ \\
\hline
\textbf{\large{K2-212b}}& &	\\
\textbf{K2 Observations}& &	\\
Transit 0           	& $2457399.641621 \pm 0.000222$ & $0.043194 \pm 0.000140$ \\
Transit 1           	& $2457409.435555 \pm 0.0002785$ & $0.044538 \pm 0.000145$ \\
Transit 2            		& $2457419.233714 \pm 0.000179 $ & $0.043518 \pm 0.000153$  \\
Transit 3					& $2457429.025492 \pm 0.000205$ & $0.044490 \pm 0.000212$ \\
Transit 4 					& $2457438.823388 \pm 0.000198$ & $0.044738 \pm 0.000179$ \\
Transit 5					& $ - $ & $ - $ \\
Transit 6					& $2457458.414117 \pm 0.000270$ & $0.043434 \pm 0.000168$ \\
Transit 7					& $2457468.206566 \pm 0.000192$ & $0.043418 \pm 0.000142$ \\
\textbf{Spitzer Observation}& &	\\
Transit 48					& $2457869.831600 \pm 0.001125$ & $0.037400 \pm 0.001551$ \\
\enddata

\end{deluxetable*}

\begin{deluxetable*}{lcc}[ht]
\renewcommand\thetable{3}
\tablecaption{Summary of Analysis Results. Below are the reported transit centers for each transit of K2-178b as observed by K2, Spitzer, and in two TESS campaigns. Times are reported in $BJD_\textrm{TDB}$.}
\tablehead{
\colhead{Transit Number} & \colhead{Central time} 
}
\startdata
\textbf{\large{K2-167b}} & &	\\
\textbf{K2 Observations} & &	\\
Transit 0           	    & $2456979.943508 \pm 0.000231$  \\
Transit 1            	    & $2456989.913294 \pm 0.000227$   \\
Transit 2				    & $2456999.926223 \pm 0.001908$  \\
Transit 3 				    & $2457009.876881 \pm 0.000966$  \\
Transit 4				    & $2457019.846160 \pm 0.000249$  \\
Transit 5				    & $2457029.839043 \pm 0.002719$  \\
\textbf{Spitzer Observation} & 	\\
Transit 47				    & $2457448.9216 \pm 0.002803$  \\
\textbf{TESS Observation} & 	\\
Transit 138					& $2458356.976 \pm 0.0032$  \\
Transit 139					& $2458366.9545 \pm 0.0032$  \\
Transit 140					& $2458376.9330 \pm 0.0032$  \\
Transit 209					& $2459065.4511 \pm 0.0032$  \\
\enddata

\end{deluxetable*}

\end{document}